# As Good as a Coin Toss

Human Detection of AI-Generated Images, Video, Audio, and Audiovisual Stimuli

DI COOKE, Department of War Studies, King's College London, London, UK
ABIGAIL EDWARDS, Center for Strategic and International Studies, Washington D.C. USA
SOPHIA BARKOFF, Center for Strategic and International Studies, Washington D.C. USA
KATHYRN KELLY, Center for Strategic and International Studies, Washington D.C. USA

One of today's principal defenses against weaponized synthetic media continues to be the ability of the targeted individual to visually or auditorily recognize AI-generated content when they encounter it. However, as the realism of synthetic media continues to rapidly improve, it is vital to have an accurate understanding of just how susceptible people currently are to potentially being misled by convincing but false AI-generated content. To ascertain this, we conducted a perceptual study with 1276 participants to assess how capable people were at distinguishing between authentic and synthetic images, audio, video, and audiovisual media. As AI-generated content is proliferating across online platforms in particular, the surveys were designed to emulate some of the ecological conditions typical of an online platform. We find that on average, people struggled to distinguish between synthetic and authentic media, with the mean detection performance close to a chance level performance of 50%. We also find that accuracy rates worsen when the stimuli contain any degree of synthetic content, features foreign languages, and the media type is a single modality. People are also less accurate at identifying synthetic images when they feature human faces, and when audiovisual stimuli have heterogeneous authenticity. Finally, we find that higher degrees of prior knowledgeability about synthetic media does not significantly impact detection accuracy rates, but age does - with older individuals performing worse than their younger counterparts. Collectively, these results highlight that it is no longer feasible to rely on people's perceptual capabilities to protect themselves against the growing threat of weaponized synthetic media, and that the need for alternative countermeasures is more critical than ever before.

### KEY INSIGHTS

- As synthetic media continues to advance in sophistication, so too does the threat posed by the growing weaponization of synthetic media, from financial fraud to the production of non-consensual intimate materials of adults and children.
- This study has found that people are now only able to distinguish between AI-generated and human-authored content 51% of the time. In addition, the presence of digital media characteristics including authenticity, modality, and subject matter content, as well as observer demographics such as multilinguism and age have also been proven to further affect an individuals' detection capabilities.
- Consequently, it has become more vital than ever before to deploy alternative countermeasures to combat the growing synthetic media misuse.

## 1 INTRODUCTION

Advancements in generative AI technology have made it easier than ever for anyone to manufacture increasingly realistic synthetic media (colloquially known as 'deepfakes') at faster speeds, larger scales, and with more customization than ever before. [15] This in turn has led to synthetic media being increasingly employed for harmful purposes, including disinformation campaigns, non-consensual pornography, financial fraud, child sexual abuse and exploitation activities, espionage, and more. [26] As of today, the principal defense to combat deceptive synthetic media depends in large part on the human observer's perceptual detection capabilities or their ability to visually or auditorily identify AI-generated content when they encounter it. [15] Yet the growing realism of synthetic media impedes this ability, heightening people's vulnerability to being misled by weaponized synthetic content. Moreover, it has been found that people overestimate how capable they are at identifying synthetic media, further exacerbating the problem. [13]

As a result, an accurate measurement of people's perceptual ability to differentiate between the real and fake is critical to effectively combat the potential and already realized harms arising from synthetic media misuse. While there are growing efforts to develop and implement alternative technical solutions, such as machine detection, watermarking, or content provenance, these methods either currently lack robustness or are not yet sufficiently widespread enough to be effective. [12, 15, 38] Similarly, despite widespread calls to deploy educational interventions such as digital media literacy campaigns or to ratify national laws requiring the labelling or prohibiting of certain AI-generated content, thus far in practice formalized efforts have been relatively limited.

We conducted a study to measure the ability of people to detect AI-generated content using their cognitive perceptual capabilities, with participants being required to classify synthetic and authentic images, audio, video, and audiovisual stimuli in a series of online surveys. In addition to reviewing general deception performance trends, our study examines how specific stimuli characteristics, including media type, authenticity, multimodality, and content subject matter, affect human detection performance. By doing so, it is possible to identify certain types of synthetic content which may be generally more difficult for people to detect, and consequently, more successful vehicles for deception. Our study also examines the impact of demographic factors – including multilingualism, age, and prior knowledgeability of synthetic media – on detection performance, revealing whether certain demographics are more vulnerable to misleading synthetic content than others.

Our survey series was designed to emulate the experience of browsing a typical online platform news feed by replicating several ecological conditions including the interface format, browsing behavior, types and subject matter of the digital media present, and the quality of synthetic stimuli being sufficiently representative of actual synthetic media being made with publicly accessible generative AI tools at the time and published online. The design details are discussed in more detail in the 'Methods' section.

As social media platforms have become a hotspot for the proliferation of AI-generated content, this approach was chosen so that study's findings would be more closely reflective of the average individual's detection performance if they were to encounter misleading synthetic media in their social media news feeds. [7] Thus far, much of human detection research has sought to assess detection performance under circumstances not necessarily reflective of how people are currently encountering synthetic media in their daily lives. Examples of this include constraining the manner of synthetic content being tested on, such as subject matter or media type, as well as employing testing conditions not present in the real world such as two-alternative forced choice methods, providing instant feedback or unlimited testing time, and informing participants what the percentage of synthetic versus authentic media will be in the test. [5, 6, 14, 16, 20] As a result, this makes it challenging to generalize many of these studies' findings for human detection performance. Comparatively, the body of literature examining human detection capabilities under conditions more typically encountered in real-life is relatively nascent – our study seeks to build on such work. For example, Joseph et al. examined the impact of common real-world video browsing conditions on synthetic video detection, such as video quality, time-limited exposure, and distraction. [8]

Our study makes several additional contributions to current literature, expanding the scope of synthetic stimuli for detection and examining several novel demographic factors. Thus far, the human detection field has predominately assessed people's ability to identify synthetic images featuring human faces. [6, 15, 16, 25, 32] This study is part of a nascent trend to examine human detection capabilities in regard to non-face images, such as urban scenes and landscapes. [1] As far as the authors are aware, this is the first study to also include synthetic images featuring animals. Furthermore, this is also the first human detection study to examine the role of language in detecting synthetic visual-based stimuli to compare detection accuracy rates between heterogeneous and fully synthetic audiovisual stimuli, to assess the impact of age on synthetic audio recognition, and to examine the effect of an individual's prior knowledgeability of synthetic media on detection performance.

Our results find that participants' overall accuracy rates for identifying synthetic content are close to a chance level performance of 50%, with minimal variation between media types, suggesting that people's visual and auditory perceptual capabilities are inadequate for reliably identifying synthetic media encountered in online platforms today. Our results also find that detection accuracy rates worsen when people are presented with stimuli featuring synthetic content as compared to authentic content, images of human faces as compared to non-human face objects, stimuli containing a single modality as compared to multimodal stimuli, and audiovisual stimuli with heterogeneous authenticity as compared to fully synthetic audiovisual stimuli. This indicates that individual content characteristics present within the stimuli do affect certain visual and auditory perceptual processes such as object, speech, and language recognition, providing potentially novel observations to not only synthetic media detection research but the human perception field as well.

We also find that demographic factors such as multilingualism and age play a role in perceptual detection capabilities. People were less accurate in correctly identifying stimuli featuring foreign languages than those featuring languages participants are fluent speakers in, and older participants performed worse than younger ones – particularly in identifying audio and audiovisual stimuli. This indicates that monolingual and older demographics may be more susceptible to being deceived by synthetic media than their multilingual and younger counterparts. Finally, we find that people's prior knowledge of synthetic media does not impact detection performance, with people who reported being unfamiliar, semi-familiar or highly familiar with synthetic media prior to taking the survey all performing similarly. This suggests either that current public knowledge of synthetic media is insufficient for meaningfully improving detection performance, or that synthetic media has become convincingly realistic enough that perceptual-based educational interventions have become inadequate.

Collectively, these results demonstrate that depending on people's perceptual detection capabilities to discern the real from the take is no longer a viable bulwark against the threats posed by synthetic media. While our findings provide useful insights on how to reduce people's susceptibility to certain digital content characteristics, such as stimuli featuring a foreign language or in a single modality, it is expected that the benefits will be relatively short term. Rather, it is expected that continued advancements in generative AI technology will eventually lead to any detection performance differences resulting from these characteristics becoming negligible in future, and that human detection performance overall will plateau. Ultimately, these findings serve to further emphasize the critical need for alternative countermeasures to more effectively combat both the potential and already realized harms arising from synthetic media, whether these measures be technical, educational, or otherwise.

## 2 RESULTS

We conducted a pre-registered perceptual survey series, requiring 1,276 participants to classify authentic and synthetic media stimuli.

**Table 1:** Logistic Regression of Participant Detection Performance

| | **β** | ***SE*** | **Wald Wald χ2** | ***df*** | ***p*** | **Odds Ratio** | **OR CI (95%)** |
|---|---|---|---|---|---|---|---|
| (Intercept) | 0.924 | 0.022 | 1767.698 | 1 | <0.001* | 2.519 | [2.808, 2.260] |
| **Authenticity** | | | | | | | |
| Synthetic Media | -1.140 | 0.012 | 8300.303 | 1 | <0.001* | 0.320 | [0.322, 0.317] |
| **Image Subject Matter** | | | | | | | |
| Human Face | -0.269 | 0.016 | 265.364 | 1 | <0.001* | 0.764 | [0.783, 0.746] |
| Language | | | | | | | |
| Foreign Language | -0.158 | 0.020 | 59.707 | 1 | <0.001* | 0.854 | [0.883, 0.826] |
| **Deepfakes Pre Knowledgeability** | | | | | | | |
| Highly Familiar | -0.035 | 0.022 | 2.332 | 1 | 0.115 | 0.966 | [1.007, 0.926] |
| Semi-Familiar | -0.005 | 0.015 | 0.095 | 1 | 0.758 | 0.995 | [1.025, 0.966] |

Likelihood ratio test = $x^2(6) = 9142.5$, $p<0.001$
Nagelkerke R2 = 0.095

### 2.1 Media Type & Authenticity

Mean detection performance across all stimuli was found to be 51.2%, close to a random chance performance of 50%. Regarding the detection performance of specific modalities, participants were the least accurate at classifying image stimuli with a mean accuracy rate of 49.4% (Figure 1). Comparatively, detection accuracy was higher for video-only stimuli and audio-only stimuli, at 50.7% and at 53.7% respectively. Participants were the most accurate when classifying audiovisual stimuli, achieving a mean accuracy of 54.5%.

It was found that the stimuli's authenticity was also a meaningful predictor for detection performance (Table 1). Across all modalities, participants were found to be significantly better at correctly identifying fully authentic stimuli (with a mean detection performance of M=64.6%) as compared to identifying stimuli which contained synthetic media (M=38.8%). When audiovisual stimuli was examined specifically, as the only multimodal media type present in the study, a unpaired t-test found that participants' detection performance worsened significantly ($t(9861)=6.3$, $p<0.001$) when classifying audiovisual stimuli containing both synthetic video and authentic audio (M=43.4%) as compared to audiovisual synthetic containing synthetic video as well as synthetic audio (M=49%). The effect size for this difference was small, with Cohen's *d* equal to 0.11.

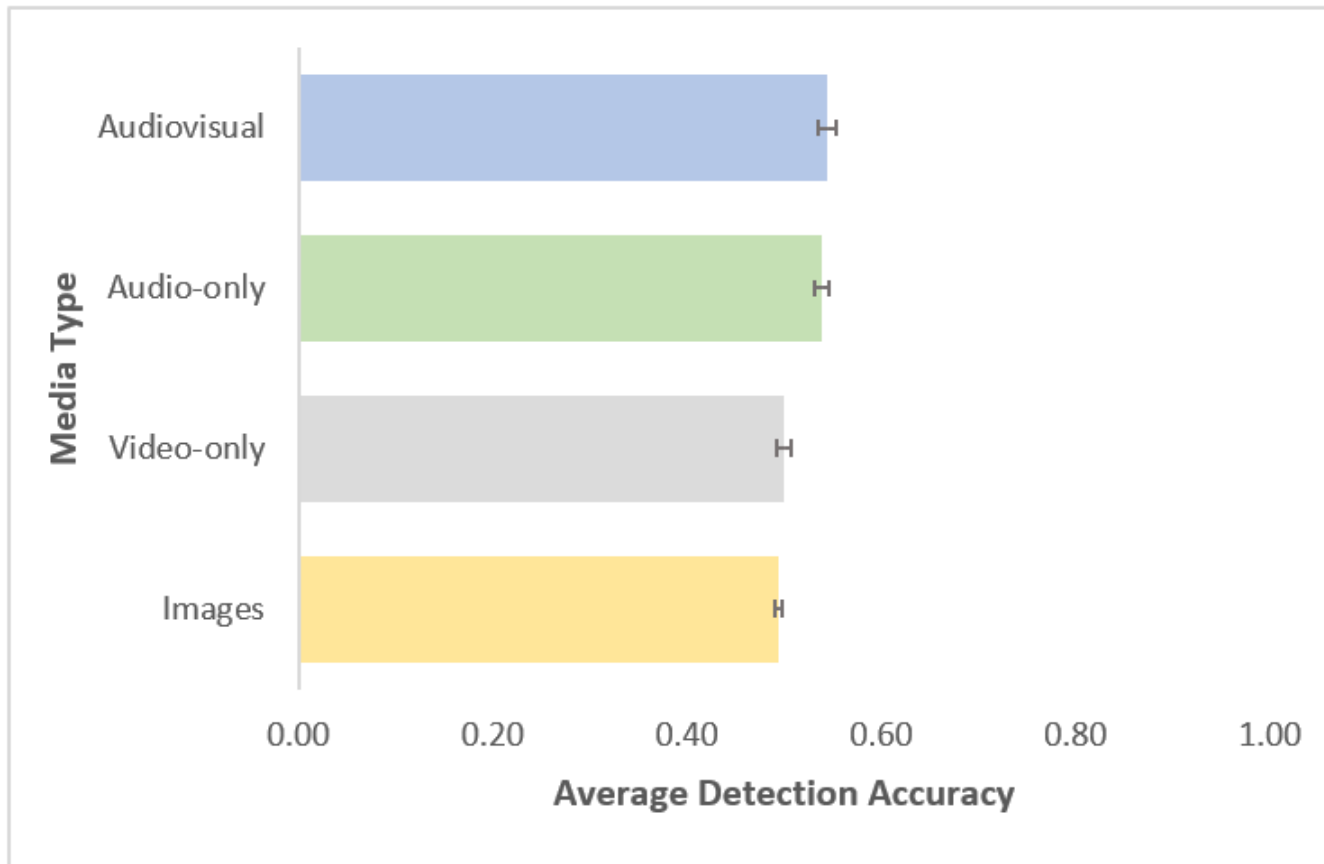

Figure 1. Mean detection accuracy by media type, with error bars representing 95% CI.

### 2.2 Human Face vs. Non-Human Face Images

Participants were also found to be significantly less accurate (Table 1) when classifying images featuring human faces (M=46.6%) as compared to images featuring non-human face objects such as animals (M=51.7%), food (M=49.9%), and landscapes (M=54.7%), shown in Figure 2. A post hoc unpaired t-test confirmed that although the effect size was small (Cohen's *d*=0.21), there continued to be a significant difference in detection performance even when controlling for model type (t(12201)=-13.83, p<0.001).

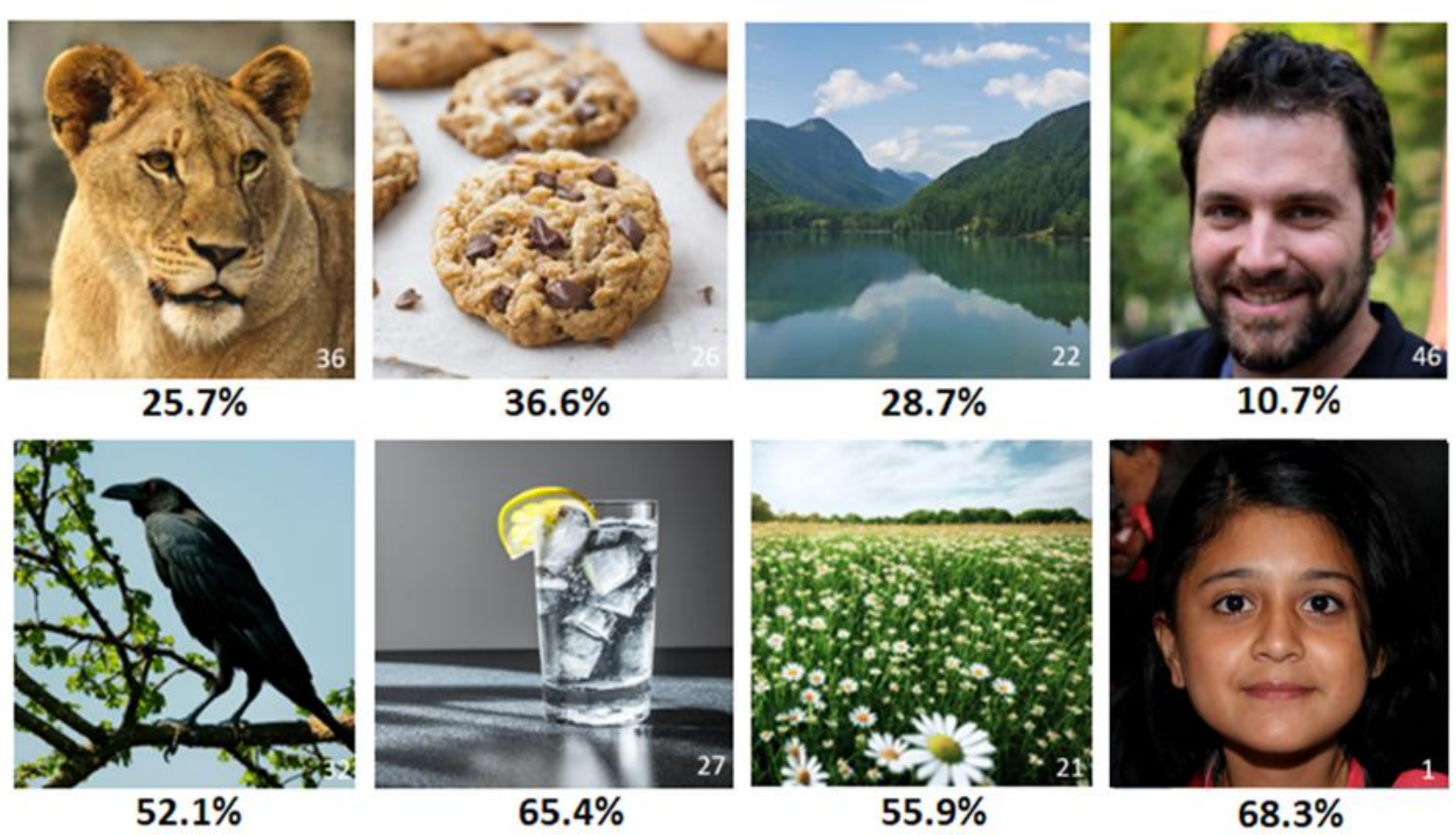

**Figure 2.** Top row: synthetic images from each category most often *misclassified* as authentic. Bottom row: synthetic images most often *correctly identified* as synthetic. Beneath each image is its corresponding mean accuracy rate.

### 2.3 Multimodality

An unpaired t-test found that the mean detection accuracy rate was significantly higher (Cohen's d=0.4, small-to-medium effect) for participants classifying multimodal audiovisual stimuli (t(49004)=5.18, p<0.001) as compared to when they classified audio-only and video-only stimuli (M=52.2%), both single modalities. 30 of the video-based stimuli were randomly presented in different formats in both of the surveys, enabling a post hoc comparison of participants' accuracy rates across each condition. In one survey, these stimuli were presented in a fully audiovisual format. In the other survey, the audio was removed and the stimuli were presented in a video-only format. Post hoc analysis found a small but meaningful effect (Cohen's d=0.11) of multimodality on participants' detection performance. An unpaired t-test found that participants were significantly more accurate (t(37517)=10.15, p<0.001) at identifying the video stimuli when the audio was included (M=55.9%) than when the same stimuli was presented in a video-only format (M=50.7%).

### 2.4 Multilinguism

It was found that detection performance significantly improved (*Appendix A: Table 1*) when the participant was required to classify visual and auditory stimuli featuring a spoken language the participant was a self-reported fluent speaker (M=54.5%) of, as opposed to classifying visual and auditory stimuli featuring a foreign language (M=51.3%). As shown in Figure 3, a series of post hoc unpaired t-tests found that detection accuracy rates continued to be significantly better when known languages were present in the stimuli as opposed to foreign languages, even when examining each modality type individually. The effect sizes across each of the three modality comparisons were small with the presence of known-languages in audiovisual stimuli having comparatively the largest effect (d=0.09), while the effect was lessened for audio-only stimuli (*d*=0.06) and was smallest for video-only stimuli (d=0.04). Participants were found to be significantly more accurate (*t*(17696)=3.947, p<0.001) at correctly identifying audio-only stimuli which featured known-languages (M=55.3%) as compared to those featuring foreign languages (M=52.3%). Detection performance was also found to be significantly better when classifying audiovisual stimuli (*t*(22640)=6.864, p<0.00) and video-only stimuli (*t*(19278)=2.83, p<0.001) featuring known languages, with mean detection accuracy rates for audiovisual (M=56.5%) and video (M=52%) stimuli featuring known languages higher as compared to mean detection accuracy rates for audiovisual (M=52%) and video stimuli featuring foreign languages (M=49.5%).

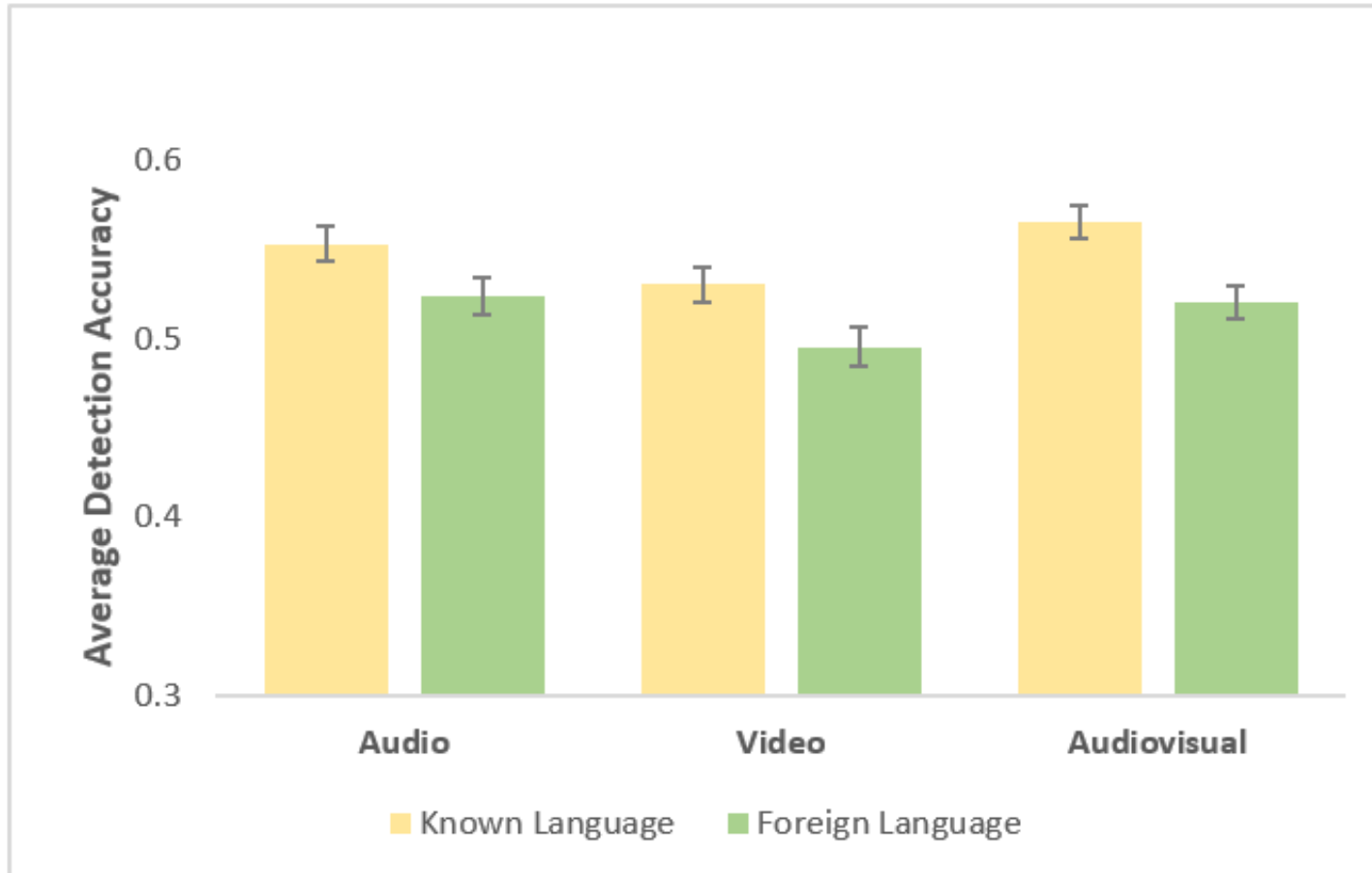

**Figure 3.** Mean detection accuracy by language familiarity between video-only clips and their audiovisual counterparts, with error bars representing 95% CI.

As seen in Figure 4, a series of additional post hoc unpaired t-tests found that when examining the 30 video-based stimuli which was present in both surveys - in either a video-only or audiovisual format - the inclusion of audio featuring a known language alongside the video resulted in a small (d=0.12) but significant improvement in participants' detection performance (t(23035)=9.17, p<0.001). Although the effect was comparatively smaller (d=0.09), detection performance also significantly improved (t(16966)=5.63, p<0.001) when audio featuring a foreign language was included alongside the video.

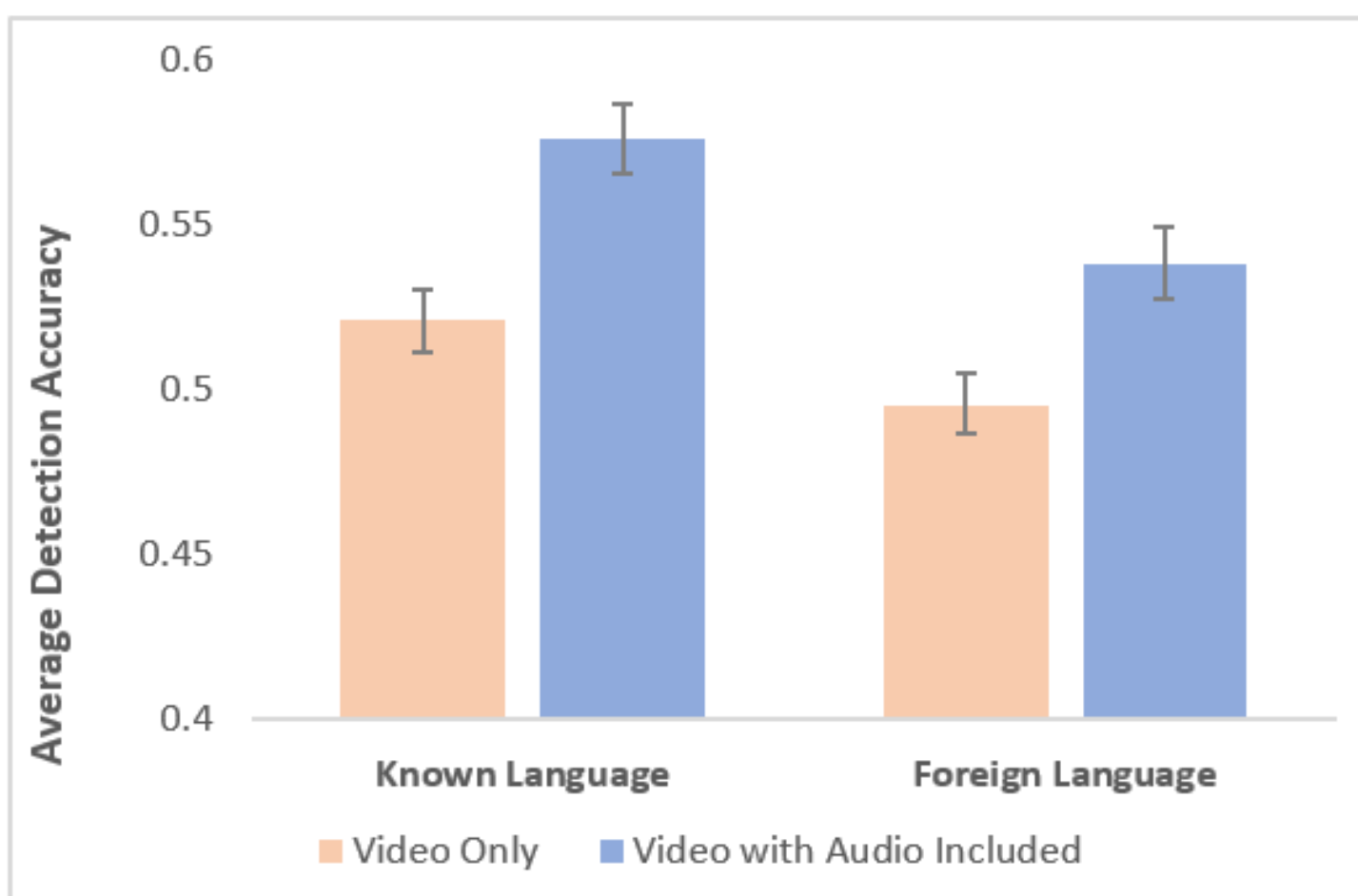

**Figure 4**. Mean detection accuracy by language familiarity between video-only clips and their audiovisual counterparts, with error bars representing 95% CI

### 2.5 Age

A post hoc linear regression analysis found age to be a significant predictor of detection performance, (β=-56.64, $p<0.001$) with older participants being significantly less accurate in classifying stimuli as compared to younger participants. [$R^2$=0.04, $F(1,1274)$=55.63, $p<0.001$] When examined by individual media type, it was found that detection performance by age decreased the greatest amount for audiovisual and audio-only stimuli, with a comparatively more constrained decrease in accuracy rates for image and video-only stimuli. (Figure 5)

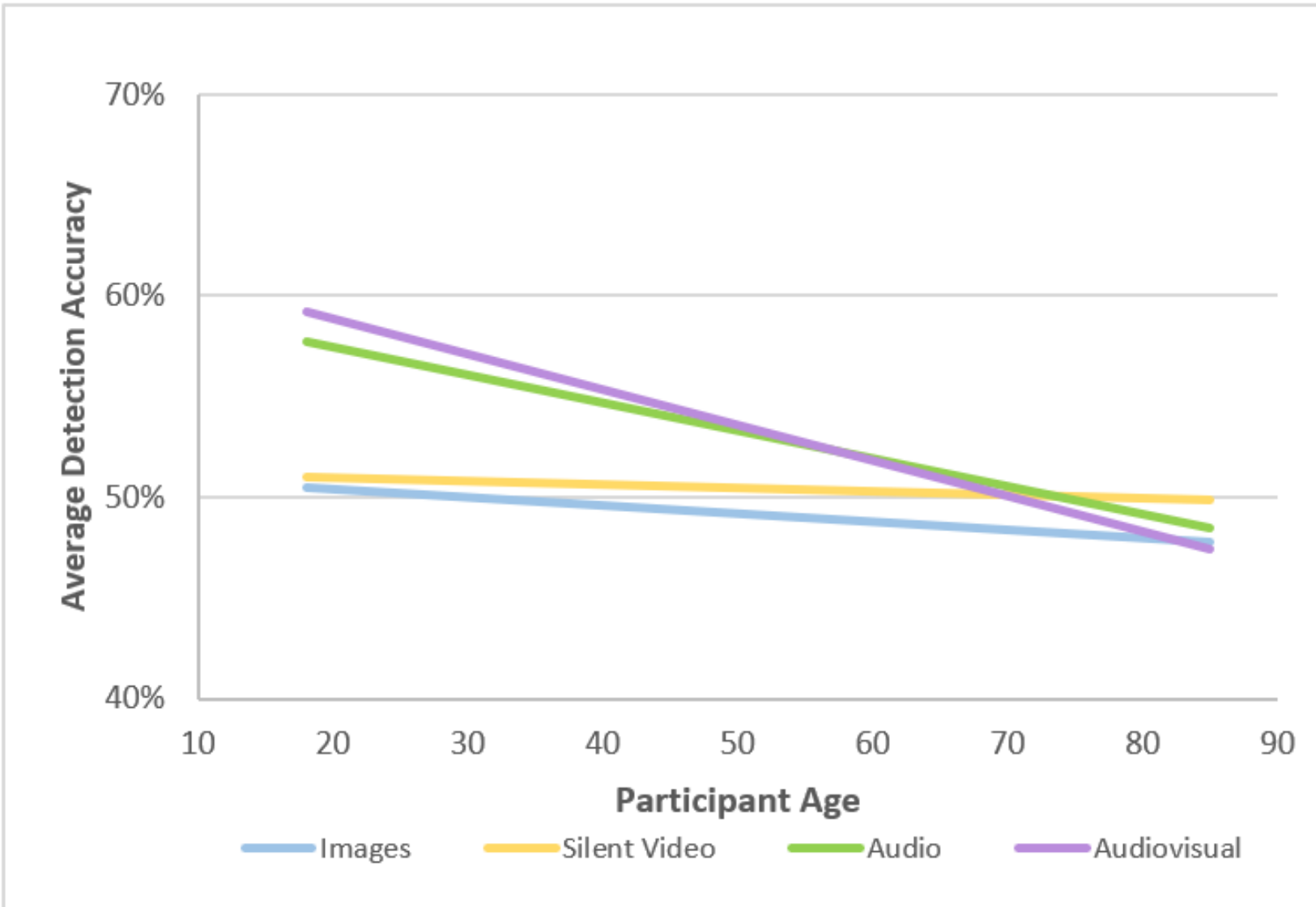

**Figure 5.** Mean detection accuracy by age, with each trendline showing detection accuracy by media type: video (yellow), image (blue), audio (green), and audiovisual (purple)

### 2.6 Prior Synthetic Media Knowledgeability

Finally, there was no significant difference in detection performance found (*Appendix A: Table 1*) between participants who self-reported being previously unfamiliar with synthetic media (M=51.1%) and those who reported having prior knowledgeability of synthetic media, either to a semi-familiar (M=51%) or highly familiar degree (M=51.9%). A post hoc one-way ANOVA did not find any significant difference in detection accuracy rates between the three respective groups [$F(2,124185)$=1.4, $p$=0.25; $\eta 2<0.01$], with Tukey's HSD test further confirming no significant difference between their means.

## 3 LIMITATIONS

This study is not without its limitations. Firstly, as advancements in generative AI technology have continued to improve the realism of synthetic content at a rapid pace current human detection performance of synthetic media at this moment may be lower than reported in this study. Furthermore, although the design of this study mimics online platform environmental conditions to replicate the manner in which individuals would likely encounter synthetic media in their day to day lives, the study does excludes several features which would typically be present in such an environment. This includes captions, or the text information accompanying a media post, metrics like likes or re-shares, and other users' comments, and does not account for the ability of participants to cross-reference, source validation, or fact-check the stimuli they encounter. Although these restrictions were done deliberately to ensure participants were solely reliant on their cognitive perceptual capabilities in identifying synthetic content, it does somewhat limit generalization of this study's findings as it does not account for people's ability to rely on other contextual cues or employ digital literacy skills to determine if a media item is real or fake. Therefore, we suggest that our study results are more ecologically valid for human detection performance on online platforms where either information is more context-limited or where there is a proven tendency of people to not critically engage with digital content. Existing literature suggests that many social media platforms are becoming increasingly such context-limited environments as a result of the growing trend for users to passively and non-critically consume digital content on these platforms, and growing reliance as a primary knowledge source of novel information such as news events. [21, 29, 35] A final limitation to be considered is that participants self-reported their degree of pre-existing synthetic media knowledge and fluency in languages, wherein participants may have over or underestimated their capabilities in comparison with other participants.

## 4 DISCUSSION

### 4.1 Social Media Platform Detection Performance

Multiple implications can be drawn from the results of this study. First, people struggled to meaningfully distinguish synthetic images, audio, video, and audiovisual stimuli from their authentic counterparts, with overall detection accuracy rates being close to a chance level performance. This is congruent with existing research to some degree, as some prior studies have recorded similar detection accuracy rates for synthetic images, [20, 32] yet studies on video, audio, and audiovisual stimuli have typically found higher detection accuracy rates. [2, 5, 6, 14, 16, 23, 25] Although study differences bar a more detailed comparison, it is hypothesized that the overall divergence in detection performance is due to the differing environmental testing conditions – which suggests that people's detection capabilities are further constrained under conditions similar to those present in browsing social media news feed. This worsening performance is consistent with Josephs et al 2023's study, which similarly found that online platform environmental factors, including divided attention and exposure length, amongst others, also had inhibitory effects on people's detection capabilities. [8] Further research might be conducted on how these and other environmental factors may also impact detection performance, both individually and cumulatively. Having a better understanding of what these environmental factors are and their effects on detection performance will provide a clearer picture of how vulnerable people may be to deceptively employ synthetic media in their daily lives. For instance, despite smartphones' growing popularity as people's primary device to access online information, it has been found that people employ lower amounts of cognitive resources when consuming digital content via smartphones. [3, 35] As a consequence, widespread smartphone use may further hinder people's abilities to detect deceptive synthetic media when consuming current events and news content via their online platform news feeds.

### 4.2 Stimuli Content Characteristics

Detection performance is found to be sensitive to certain stimuli content characteristics, which suggests that these individual factors may affect people's detection capabilities because of how they impact their cognitive perceptual processes.

**Authenticity**: People are more accurate at identifying authentic content rather than synthetic, revealing a bias towards classifying all digital content as being real. This is consistent with previous research which also found people had a similar inclination to classify video-only stimuli as being authentic. [8, 13] **Modality**: People are also less accurate at detecting synthetic audio-only and video-only stimuli as compared to audiovisual, which suggests the inclusion of additional modalities to stimuli improves people's perceptual ability to distinguish between authentic and synthetic media. This is consistent with existing research such as Groh et al 2023's study, which found the addition of audio to videos of real and fake political speeches improved detection performance. [6] This may be due to the multimodal nature of speech perception, as speech perception research has found that people rely on both visual and auditory perceptual cues to comprehend spoken words. [24] Therefore, synthetic audiovisual stimuli may be less difficult to detect than its monomodal counterparts because people are able to leverage both the auditory and visual information present to better identify observable AI-generated artifacts present in the stimuli. In addition, detection accuracy is found to be higher when identifying fully synthetic audiovisual stimuli, as compared to heterogeneous synthetic audiovisual stimuli where the video contains synthetic content while the audio is authentic. This suggests that people may be better at identifying fully synthetic audiovisual stimuli than heterogeneous due to the higher potential presence of observable artifacts in both the audio and visual modalities which they are able to leverage to determine authenticity, as opposed to potential observable artifacts only being present in only one modality as would be the case with heterogeneous synthetic audiovisual stimuli. **Image Content**: Meanwhile, people are worse at detecting synthetic images featuring human faces as compared to synthetic images featuring animals, food, and landscapes, indicating they found synthetic images featuring human faces to be more convincing than similarly realistic looking images featuring non-human face images. This may be due to the specialized human visual perceptual process which occurs when people observe faces as opposed to non-face objects. Visual perception research has established that whereas humans recognize faces as a perceptual whole, non-face objects are conversely identified by their distinct individual components. [11, 30] Therefore, the more gestalt face recognition process which takes place once an image is recognized to feature a face may lead people to be less sensitive towards identifying observable AI-generated artifacts than they would if employing the more fragmented object recognition process which occurs when observing non-face objects.

### 4.3 Multilingualism

People are also more accurate in detecting synthetic audio-only, video-only, and audiovisual stimuli featuring languages the observer is a fluent speaker of compared to stimuli featuring foreign languages, suggesting that language familiarity plays a significant role in people's visual and auditory perceptual detection capabilities. This is consistent with existing synthetic audio detection research such as Müller et al's 2023 study, which found native English speakers to be better at detecting English synthetic audio clips than non-native speakers. [23] Detection performance may be higher when known languages are present in visual and auditory stimuli due to the observer being more familiar with the visual and auditory language information available, making them more sensitive to observable artifacts. [18] In addition, when comparing detection performance of video-only stimuli as compared to their audiovisual counterparts (identical video clips but with audio included) it was found that people's detection accuracy improved to a greater degree between video-only and audiovisual stimuli featuring known languages versus those featuring foreign languages. This suggests that people are more sensitive to auditory perceptual cues in audiovisual stimuli featuring familiar language as opposed to foreign ones. This is congruent with established language perception research, which has found that people weigh auditory information more highly when observing known languages being spoken, while they rely on visual perceptual cues to a greater degree when observing foreign languages. [27, 36] Therefore, it may be that the inclusion of audio facilitated people's sensitivity to synthetic visual content to a greater degree for stimuli featuring familiar languages than foreign languages because of the greater weight given to auditory perceptual cues when observing familiar languages as part of the language perceptual process.

Understanding how individual stimuli characteristics such as the ones examined in this study may inhibit or facilitate people's perceptual detection capabilities is valuable as it provides more accurate predictions of how susceptible people potentially are to different types of synthetic content they may be exposed to. In addition, it identifies demographics that may be more vulnerable to being duped by synthetic media than others, such as monolingual speakers potentially being less sensitive to visual and auditory AI-generated artifacts presented in foreign languages than multilingual speakers. In turn, these insights can inform the development of more effective countermeasures to better mitigate people's susceptibility to synthetic media, such as specific content moderation tactics. For instance, although the default for many social media platforms is to have audiovisual stimuli muted when being played, detection performance of synthetic media is likely to be improved if the audio is not muted when the video automatically plays. Therefore, a useful content moderation policy to adopt may be to not automatically mute videos deemed to be at higher risk of containing deceptive synthetic content, such those featuring political content in the run up to government elections. Future research would be useful to identify additional actionable countermeasures as well as further explore the effects of these and other stimuli characteristics on detection performance, such as determining whether the difference in detection performance between faces and non-face objects is also found in identifying synthetic video and audiovisual stimuli. Regardless, as generative AI technology continues to improve, periodic reassessment of these stimuli content characteristics would be beneficial to determine whether they continue to have the same effect.

### 4.4 Age

People's age significantly affected detection performance, with older people being less accurate in identifying synthetic media than younger people across all media types. The relationship between age and detection performance is consistent with previous research, which similarly found older people to be less accurate in classifying audiovisual and audio stimuli, respectively. [2, 16] This suggests that as people age, they become less sensitive to perceptual cues such as observable artifacts present within synthetic media. We speculate that this may be due to the widespread visual and auditory perceptual degradation which occurs with aging. [4] Detection performance in relation to older age declined the most for stimuli containing audio components, which may be a result of hearing loss being more prevalent than visual impairment in older demographics, along with the lower frequency of people using treatments for improving their hearing in comparison to their sight. [4, 28] Audiovisual stimuli, which younger people were the most accurate at classifying, was the media type least often correctly classified by older people. This may be due to how age-related auditory and visual degradation impacts the multisensory perceptual process. Existing speech perception research finds that older individuals depend more on multisensory information to compensate for degradation of unisensory perception, such as relying more on adjacent visual cues to improve their comprehension of auditory information. [22] This higher reliance on the integration of perceptual information has led to older individuals being more likely to be susceptible to illusionary effects, such as purposefully mismatched visual and auditory information, than their younger counterparts. [22] Similarly, this may mean older individuals are less sensitive to artifacts present in audiovisual stimuli because they are more reliant on utilizing their multisensory perception capabilities than unisensory when observing audiovisual content.

Altogether, this indicates that age increases susceptibility to being misled by synthetic media, with a greater vulnerability to audio and audiovisual synthetic content than visual. This insight suggests the potential need for age-related content moderation policies to better improve older digital consumer's resilience against being tricked by deceptive synthetic media, or for the heighted need of educational interventions tailored to this higher risk demographic. This is especially important considering older individuals are increasingly becoming the targets of synthetic media enhanced scams. [22] Future research on how age-related limitations may be mitigated would be useful, such assessing whether corrective devices such as glasses or hearing aids meaningfully improve perceptual detection performance

### 4.5 Prior Synthetic Media Knowledgeability

People's prior knowledgeability of synthetic media did not affect detection performance, with people who reported being highly familiar with synthetic media performing similarly to those who reported being less familiar or unfamiliar. As this study did not test participants on their synthetic media knowledge, rather asked them to self-report, this suggests one of two possible causes. The first is that synthetic media has become convincingly realistic to the degree where increased familiarity with synthetic media does not meaningfully improve people's perceptual detection capabilities. Alternatively, it may be that current public knowledge of synthetic media, even at comparatively higher levels, does not sufficiently educate people on effective perception-based detection methods. Existing research suggests the truth is currently somewhere in the middle, as some studies have found that increased exposure to synthetic content and providing immediate feedback or prior training has improved detection performance, while others have found that not to be the case. [5, 6, 14, 16] However, as synthetic media continues to improve in realism, it is expected that eventually perceptual-based educational interventions will become less effective. Nonetheless, further work to clarify in what contexts perceptual-based interventions are able to improve detection performance will be beneficial for developing more effective educational interventions to reduce people's current vulnerability to deceptive synthetic content. In addition, research into the impact of non-perception based educational interventions, such teaching critical analysis skills like fact-checking or cross-referencing, on detection performance would also be valuable.

## 5 CONCLUSION

The results of our study demonstrate that people's perceptual detection capabilities are no longer a suitable defense against deceptive synthetic media. The ability to create synthetic content which is convincingly realistic has now become available for any member of the public to use, including those with harmful intent. Our findings underscore the critical importance of developing and deploying robust countermeasures which are not reliant on human perceptual detection capabilities. This includes increasing investment and research for technical solutions such as machine detection, watermarking or cryptographic signatures, as well as the wider adoption of other techniques like content provenance or hashing databases. It also highlights the need to pursue widespread educational interventions such as digital media literacy campaigns to better equip people with the skills and knowledge to identify false content in other ways, such as critical analysis techniques like cross-referencing and fact checking.

Importantly, our study also identifies several stimuli content characteristics and observer demographic factors that have inhibitory or facilitatory effects on people's perceptual detection capabilities. These findings provide useful insights for informing immediate actionable countermeasures which could be taken in the short term to reduce people's vulnerability to more convincingly deceptive content, such as specific content moderation policies for online platforms. Nevertheless, as synthetic media outputs continue to progress realism, it is anticipated that perceptual-based countermeasures will eventually plateau, requiring alternative solutions over the long term. Regardless, further work in this space is vital to improve our understanding of and monitor the limitations of human perceptual deception capabilities to better identify and improve societal resilience against the dangers posed by synthetic media.

## 6 METHODS

### 6.1 Overall Study Design

Our pre-registered mixed design study was divided across two online surveys, with participants only able to take one of the two surveys. This was done to reduce the risk of participant exhaustion and to eliminate carryover effects. All participants gave fully informed consent prior to taking part in the study. Participants were required to use a computer to take the survey and were provided with an introduction to the study's purpose and explanation of what synthetic media was.

Prior to beginning the survey, participants were required to report what other languages apart from English they were fluent speakers in. To control for the linguistic diversity of these participants, their answers about stimuli which featured the non-English language the participant had reported being fluent speakers in were individually re-categorized under the 'known language' dataset. Participants were also asked to select what level of preexisting knowledge of synthetic media appropriately represented them, with the levels being either: 1) Highly Familiar ("*I have a comprehensive understanding what kind of deepfakes can be made, and have encountered many examples before*"), 2) Semi-Familiar ("*I have a general understanding of deepfakes, and have encountered a couple of examples before*"), or 3) Unfamiliar ("*I've never heard of deepfakes before, or I recognize the term but don't know much about them, and have not encountered examples yet that I can recall*.")

Participants were asked to classify the images, audio-only, and video-only stimuli as either being authentic or as containing synthetic media. (See *Appendix B*) They were asked to classify fully audiovisual stimuli as either being entirely authentic, containing authentic audio but synthetic visual content, containing synthetic audio but authentic visual content, or both audio and visuals containing synthetic content. The stimuli were presented in randomized order and the audio, video, and audiovisual clips were all repayable. Two attention checks each were presented at random within Survey 1 subsections 1 and 2, while two attention checks were presented at random in Survey 2. Participants who failed to pass both attention checks within each Survey 1 subsection or within Survey 2 had their results removed from that section respectively. Participants who dropped out part way through a section or reported having issues with playing the audio, video, or audiovisual clips, had their results removed from those sections as well.

In subsection 1 of Survey 1, 663 participants classified 96 image stimuli and passed both attention checks. Image stimuli were presented in a separate subsection from the other media types so that if participants had difficulties playing the audio, video, and audiovisual clips, requiring their results to be removed, this would not also require removing their image stimuli results as well. In subsection 2 of Survey 1, 604 participants from the same group of participants classified 48 stimuli including 14 audio-only, 16 video-only, and 18 audiovisual clips, passing both attention checks. In Survey 2, 614 novel participants classified a different set of 16 audio-only, 14 video-only, and 18 audiovisual stimuli, passing both attention checks. The video-only stimuli presented in Survey 2 contained the same visual content as the audiovisual stimuli from Survey 1 but with the audio content removed, with the reverse for the Survey 1 video-only and Survey 2 audiovisual stimuli.

### 6.2 Analysis Approach

Due to the diversity of variables being examined, we ran a binomial multiple logistic regression (*Appendix A: Table 1*) so that the concurrent effect of these several factors on detection performance could be studied together, ensuring that the independent effect of each variable could be evaluated individually while controlling for the effect of the other variables simultaneously. The variables analyzed in the model include the presence of synthetic content in the stimuli, the presence of human face or non-human face objects in image stimuli, whether stimuli featured languages that were known or foreign to the participant, and the participant's self-reported level of pre-existing knowledgeability on synthetic media. To further interrogate the variables' individual effects in greater detail, a series of ANOVA and t-tests were also conducted. Finally, post-hoc ANOVA and t-tests were conducted on variables added post-registration, including stimuli multimodality and participant age. Effect sizes were calculated using Cohen's D. As several tests were performed on the same samples, to reduce the risk of type 1 error the study's p-value was adjusted to $p=0.0031$ ($\alpha/16$) using Bonferroni correction.

### 6.3 Replicating a Social Media News Feed

The survey was designed to emulate the experience of browsing an online platform by mimicking typical platform characteristics. The first of this is the survey's interface format, where participants were required to progress through the survey scrolling vertically with stimuli being presented to them sequentially, much like how a digital consumer would scroll through their social media news feed and consume information post-by-post. (*Appendix B*) Next, participants were asked at the beginning to progress through the survey at a pace reflective of how they would browse through an online platform's newsfeed, so as to emulate the time-limited exposure people typically experienced when consuming media via browsing. [17, 34] A self-regulated approach to pacing was decided to be appropriate for this study, as imposing hard time limits would risk insufficiently accounting for the variety of factors which existing research has shown to affect browsing speed, including media type, subject matter, emotional valence, and personality. [17, 34, 37] Although there were no hard time restraints, with participants having the freedom to decide how much time they spent on each question, signposts were included at regular intervals to remind them to continue to take the survey at their typical browsing speed appeared when participants completed 25%, 50%, and 75% of the survey.

In addition, the type of media items present in the survey were deliberately representative of the most common digital media on online platforms, including images of people, food, landscapes, and animals, as well as user generated social media videos, film scenes, news segments, music videos, vlogs, podcasts and audiobooks clips, and radio segments. [31] Also, audio content included many of the most widely spoken languages found online. [33] To ensure the standard of the synthetic stimuli used in the survey was reflective of the average quality of synthetic content being published widely on online platforms, all the synthetic media items were sourced only from either commercially available generative AI products and services or from open-source software accessible to the public. Meanwhile, participants were not provided with any feedback, nor were they advised what the proportion of synthetic to authentic content there would be, as while these are common design characteristics present in previous human detection studies, such information sharing does not occur when individuals encounter deceptive synthetic content on social media platforms. [5, 6, 14, 16, 20]

### 6.4 Participants

A total of 124,187 observations were collected and retained from 1,276 participants for this pre-registered study. All participants were fluent English speakers and were North American residents. The sex and age demographic distributions were representative of US demographics: Survey 1 being 45% Female, 8% Unreported; 19% 18-29 years old, 62% 30-64 years old, 19% 65+ years old and Survey 2 being 49% Female, 2% Unreported; 19% 18-29 years old, 61% 30-64 years old, 20% 65+ years old. 75 participants reported being fluent speakers in one other language, including Spanish (35 participants), Mandarin (12), Portuguese (4), German (4), Dutch (3), Hindi (3), Italian (3), Korean (3), French, (2), Hebrew (1), Lithuanian (1), Romanian (1), Russian (1), Turkish (1), Vietnamese (1). Participants were recruited from the research survey platform Prolific and were paid a pro-rata rate of $11.4 per hour.

### 6.5 Stimuli

A total of 194 stimuli were presented to participants across both surveys, consisting of authentic and synthetic images, audio-only, video-only, and audiovisual clips. Image, audio-only, and video-only stimuli were each 50% synthetic and 50% authentic. Of the audiovisual stimuli, 15 clips were fully authentic, 15 contained synthetic video and authentic audio, and 8 were fully synthetic. In all synthetic stimuli, AI-generated content was prominently featured. Audio content featured a variety of languages including English, Mandarin, Spanish, Hindi, Turkish, Russian, Portuguese, French, German, Hebrew, Swedish, Japanese, and Korean. All synthetic content was sourced from commercially available generative AI products and services or popular open-source software prior to 2024, leveraging a variety of model types employed at the time, including GANs, diffusion models, and generative autoencoders. Synthetic stimuli were either produced for the purposes of this study or were collected from previously published digital content. Authentic stimuli were collected from commensurate digital content on multiple online platforms and manually curated to match the synthetic stimuli in terms of subject matter, content types, and quality. Synthetic stimuli collected from pre-existing digital content was verified by confirming the source content contained observable AI-generated artifacts. The legitimacy of authentic stimuli was verified by being published as human-created content by a credible source.

To ensure participants relied predominantly on their perceptual detection capabilities to ascertain the authenticity of a media item, rather than on alternative contextual sources of information, extraneous cues within the content were minimized or excluded, such as unique backgrounds or memorable contextual information. However, as contextual content could not be fully removed from all stimuli, diverse subject matters were purposely included, such as clips from foreign or less well-known films, interviews, and newscasts or user-generated media from social media to further reduce possible participant reliance on contextual cues for detection.

**Images**: 96 images each containing single subject matter were used, with of them prominently and separately featuring 48 human faces, 26 animals, 12 landscapes, and 10 of food. Synthetic images of human faces were sourced from multiple publicly available general adversarial network (GAN) datasets and were manually curated to be diverse across sex, race, and age (Male-presenting, Female-presenting; Caucasian, Black, East Asian, South Asian; Child, Young Adult, Middle Aged Adult, Elderly Adult) [9, 10]. (*Appendix C: 16, 19, 21, 25, 26*) Authentic human faces were sourced from the FFHQ dataset and were matched against the synthetic dataset in terms of race, sex, and age. [39] Synthetic non-face images were sourced from open source or publicly available GAN and latent diffusion (LD) datasets or directly from models. (*Appendix C: 10, 17, 19, 20 25, 26)* **Audio-Only**: 30 audio stimuli consisting of 5- to 6-second-long clips were used, each featuring a human voice speaking clearly in a single language. Synthetic audio clips were specifically generated for this study or collected from pre-existing outputs. (*Appendix C: 5, 7-9, 23, 24*) While exact model information was not available for all stimuli, audio manipulation techniques employed to produce the clips included text-to-voice (TTV), voice cloning, and voice masking. [12] **Audiovisual & Video-Only**: 38 audiovisual and 30 video stimuli in both vertical and horizontal formats were used. Each clip was between 5 to 6 seconds long and featured a human speaking clearly in a single language. In all the stimuli the human speaking features prominently and their face is fully visible the entire clip. Synthetic stimuli were sourced from published outputs produced from open-source and commercially available generative AI software and services, employing models such as generative autoencoders. (*Appendix C: 1-6, 11, 12, 14-16, 24, 24*) While exact model information was not available for all stimuli, video manipulation techniques employed to produce the clips included face swapping, head generation, and lip syncing, as well as audio manipulation techniques such as voice cloning, TTV, and voice masking. [19]

## 7 DATA

The study pre-registration can be found at https://osf.io/fnhr3. Study datasets are not yet publicly available but can be made available upon request in support of reviewing this submission. We declare no competing interests.

## ACKNOWLEDGMENTS

We would like to thank Gamin Kim, Ike Barrash, and Daniel Pycock for their contributions in data analysis and formatting, as well as Alexis Day for her contributions to the survey's design and development.